\documentclass[12pt,a4paper]{article}
\usepackage{amsmath}
\usepackage{amssymb}
\usepackage{enumerate}
\usepackage{amsfonts}
\usepackage[latin1]{inputenc}
\begin{document}

%
\newcounter{saveeqn}
\newcommand{\alpheqn}{\setcounter{saveeqn}{\value{equation}}%
\setcounter{equation}{0}%
\addtocounter{saveeqn}{1}
\renewcommand{\theequation}{\mbox{\arabic{saveeqn}\alph{equation}}}}
\newcommand{\reseteqn}{\setcounter{equation}{\value{saveeqn}}%
\renewcommand{\theequation}{\arabic{equation}}}
\begin{center}
{\bf ``Polyakov D1 Brane Action On the Light-Front'' \footnote{ ``Invited Contributed Talk'' delivered at the International Conference On "Light Cone 2008: Relativistic  Nuclear and Particle Physics (LC2008)'',  Mulhouse, France, July 07-11, 2008.} }
\\[35mm]

D.S. Kulshreshtha \\

\vskip1cm

Department of Physics and Astrophysics, \\
 University of Delhi, Delhi-110007, India. \\
 Email: dskulsh@gmail.com \\

\end{center}
\vspace{4cm}

\begin{abstract}
Conformally gauge-fixed Polyakov D1 brane action is seen to be a constrained system in the sense of Dirac when it is considered on the light-front in contrast to the case when it is consdired in the instant-form. The model is quantized using the standard constraint quantization techniques on the light-front.
\end{abstract}

\newpage

\section{\bf Conformally Gauge-Fixed Polyakov D1 Brane \\Action}
Conformally gauge-fixed Polyakov D1 brane action \cite{1,2} is quantized on the light-front using the equal light-cone (LC) world-sheet (WS) time framework on the hyperplanes of the LF defined by the LC-WS time $\sigma^{+} = (\tau + \sigma) = $ constant. The Polyakov D1 brane action which describes the propagation of a D1 brane in a d-dimensional curved background $h_{\alpha \beta}$ (with d = 10 for the fermionic and d = 26 for bosonic D1 brane) is defined by \cite{1,2}:

\alpheqn
\begin{eqnarray}
\tilde S &=& \int \tilde{\cal L} d^{2} \sigma \\ 
\label{1a}
\tilde{\cal L} &=&  \biggl[ - \frac{T}{2} {\sqrt{-h}} h^{\alpha\beta} G_{\alpha\beta} \biggr] \quad ; \quad  h = \det (h_{\alpha\beta}) \\
\label{1b}
G_{\alpha\beta} &=& {\partial}_{\alpha} X^{\mu} {\partial}_{\beta} X^{\nu} {\eta}_{\mu\nu}; ~ {\eta}_{\mu\nu} = {\rm diag} (-1,+1,...,+1) \\
\label{1c}
\mu,\nu &=& 0,1, i ; ~  i = 2,3,....,(d-1) \quad;\quad \alpha,\beta = 0,1
\label{1d} 
\end{eqnarray}
\reseteqn

Here $\sigma^{\alpha} \equiv (\tau,\sigma)$ are the two parameters describing the worldsheet (WS). The overdots and primes would denote the derivatives with respect to $\tau$ and $\sigma$. $T$ is the string tension. $G_{\alpha\beta}$ is the induced metric on the WS and $X^{\mu}(\tau, \sigma)$ are the maps of the WS into the $d$-dimensional Minkowski space and describe the strings evolution in space-time \cite{1,2}. $h_{\alpha\beta}$ are the auxiliary fields (which turn out to be proportional to the metric tensor ${\eta}_{\alpha\beta}$ of the two-dimensional surface swept out by the string). One can think of ${\tilde S} $ as the action describing $d$ massless scalar fields $X^{\mu}$ in two dimensions moving on a curved background $h_{\alpha\beta}$. Also because the metric components $h_{\alpha\beta}$ are varied in the above equation, the 2-dimensional gravitational field $h_{\alpha\beta}$ is treated not as a given background field, but rather as an adjustable quantity coupled to the scalar fields \cite{1}. The action ${\tilde S} $ has the well-known three local gauge symmetries given by the 2-dimensional WS reparametrization invariance (WSRI) and the Weyl invariance (WI) \cite{1,2}. We could  use the three local gauge symmetries of the theory to choose $h_{\alpha\beta}$ to be of a particular form \cite{1,2}:

\alpheqn
\begin{eqnarray}
h_{\alpha\beta} &=& {\eta}_{\alpha\beta} = 
\left( \begin{array}{ll} -1 & ~~0 \\ ~~0 & +1 \end{array} \right)  \\
{\sqrt{-h}} &=& {\sqrt{-\det(h_{\alpha\beta})} } = +1
\label{2}
\end{eqnarray}
\reseteqn

This is the so-called conformal gauge (CG) and the action ${\tilde S} $ in this CG becomes:

\alpheqn
\begin{eqnarray}
S^{N}  &=&  \int {\cal L}^{N}  d^{2} \sigma \\
\label{3a}
{\cal L}^{N}  &=& (-T/2) {\sqrt{-h}} h^{\alpha\beta} G_{\alpha\beta} \\
\label{3b}
 &=& (-T/2) {\partial}^{\beta} X^{\mu} {\partial}_{\beta} X_{\mu} \\
\label{3c}
\mu,\nu &=& 0,1, i ; ~  i = 2,3,....,(d-1) \quad;\quad \alpha,\beta = 0,1
\label{3d} 
\end{eqnarray} 
\reseteqn

The canonical momenta conjugate to $X^{\mu}$ obtained from the above equation is seen to be expressible and therefore the system is unconstrained in the sense of Dirac\cite{4} and the quantization of the system is therefore trivial. The nonvanishing equal WS-time (EWST) commutation relations for the theory are given by \cite{1,2}:

\begin{equation}
[ X^{\mu}(\sigma, \tau)~ ,~ P_{\nu} (\sigma' , \tau)] = i {\delta}^{\mu}_{\nu}  \delta(\sigma - \sigma')
\label{4}
\end{equation}

where $\delta( \sigma - \sigma' )$ is the Dirac distribution function.

\section{Light-Front Quantization }

For the LFQ of the theory we use the three local gauge symmetries of the theory to choose $h_{\alpha\beta}$ to be of a particular form as follows:

\alpheqn
\begin{eqnarray}
h_{\alpha\beta} &:=& {\eta}_{\alpha\beta} = 
\left( \begin{array}{ll}  ~~~~0 & -1/2 \\ -1/2 & ~~~~0 \end{array} \right) \\
{\sqrt{-h}} &=& {\sqrt{- \det(h_{\alpha\beta)}} } = + 1/2  \\
h^{\alpha\beta} &:=& {\eta}^{\alpha\beta} = \left( \begin{array}{ll}  ~~0 & -2 \\ -2 & ~~0 \end{array} \right)
\label{5}
\end{eqnarray}
\reseteqn

This is the so-called conformal gauge (CG) in the LFQ of the theory. In this LC formulation, we  use the LC variables defined by \cite{1,2}:

\begin{equation}
{\sigma}^{\pm} := (\tau \pm \sigma) \quad {\rm and} \quad X^{\pm} := (X^{0} \pm X^{1})/ {\sqrt 2}
\label{6}
\end{equation}

The action ${\tilde S} $ in the above CG, in the LF quantization reads:

\alpheqn
\begin{eqnarray}
S^{N} &=& \int {\cal L}^{N} d {\sigma}^{+} d {\sigma}^{-} \\
\label{7a}
{\cal L}^{N} &=& (-T/2) {\partial}^{\beta} X^{\mu} {\partial}_{\beta} X_{\mu} \\ \label{7b}
&=& \biggl[\frac{-T}{2} \biggr] \biggl[ ({\partial}_{+} X^{+}) ({\partial}_{-}  X^{-} ) + ({\partial}_{+} X^{-} )({\partial}_{-} X^{+} ) + ({\partial}_{+} X^{i} ) ({\partial}_{-} X^{i} )\biggr]  \\
\label{7c}
\mu,\nu &=& + , - , i \quad ;\quad i = 2,3,... (d-1) \quad;\quad \alpha , \beta = + , -
\label{7d}
\end{eqnarray} 
\reseteqn

Now onwards we study the Polyakov D1 brane LC action $S^{N} $  defined by the above equation. This theory is seen to possess 26 primary constraints\cite{2,3,4}:

\alpheqn
\begin{eqnarray}
{\chi}_{1} & = & (P^{+} + \frac{T}{2} {\partial}_{-} X^{+}) \approx 0 \\
\label{8a}
{\chi}_{2} & = &  (P^{-} + \frac{T}{2} {\partial}_{-} X^{-}) \approx 0 \\
\label{8b}
{\chi}_{i} & = & (P_{i} + \frac{T}{2}  {\partial}_{-} X^{i} ) \approx 0 
 \quad ; \quad  i = 2,3,.....,(d-1).
\label{8c}
\end{eqnarray}
\reseteqn

The canonical Hamiltonian density corresponding to ${\cal L}^{N} $ is seen to vanish weekly in the sense of Dirac\cite{4}. After including these primary constraints in the canonical Hamiltonian density ${\cal H}^{N}_{c}$ with the help of Lagrange multiplier fields $u, v$ and $w_{i}$, which are dynamical, the total Hamiltonian density ${\cal H}^{N}_{T} $ could be written as

\begin{equation}
{\cal H}^{N}_{T}  = \biggl[ u(P^{+} + \frac{T}{2}  {\partial}_{-} X^{+}) + v( P^{-} + \frac{T}{2} {\partial}_{-} X^{-} ) + w_{i} (P_{i} + \frac{T}{2} {\partial}_{-} X^{i}) \biggr]
\label{9}
\end{equation}

The Hamilton's equations obtained from the total Hamiltonian  are the equations of motion of the theory that preserve the constraints of the theory in the course of time. Demanding that the primary constraints ${\chi}_{1}, {\chi}_{2}$ and ${\chi}_{i}$ be preserved in the course of time one does not get any secondary constraints. The theory is thus seen to possess only 26 constraints ${\chi}_{1}, {\chi}_{2}$ and ${\chi}_{i}$. The Hamilton's equations obtained from the above total Hamiltonian describe the correct dynamics of the system. Now, following the standard Dirac quantization procedure in the Hamiltonian formulation \cite{4}, the nonvanishing equal LC world-sheet-time (ELCWST) commutators of the theory described by the Polyakov D1 brane LC action $S^{N} $ could be obtained after a lengthy but straight forward calculation and are omitted here for the sake of brevity\cite{3,4,5}. In the path integral formulation, the transition to the quantum theory, is, however, made by writing the vacuum to vacuum transition amplitude called the generating functional $Z [ J_{k} ]$ of the theory in the presence of external sources $J_{k}$ which is obtained for the present theory  as follows \cite{2,3} :

\begin{equation}
Z [J_{k}] := \int[d\mu] \exp \biggl[ i \int d {\sigma}^{+} d {\sigma}^{-}\biggl[  (- T/2) [ u({\partial}_{-} X^{+}) + v({\partial}_{-} X^{-} ) + w_{i}({\partial}_{-} X^{i} ) ]  + J_{k} {\Phi}^{k}  \biggr]
\label{10}
\end{equation}

where the phase space variables of the theory are $ {\Phi}^{k} \equiv (X^{+},X^{-}, X^{i},u,v,w_{i} )$ with the corresponding respective canonical conjugate momenta: ${\Pi}_{k} \equiv (P^{-},P^{+},P_{i},p_{u},p_{v}, {p}_{w_{i}} )$. The functional measure $[d\mu]$ of the generating functional $Z [J_{k}]$ is obtained as \cite{2,3}:

\begin{eqnarray}
[d\mu] &=& [T {\partial}_{-} \delta( {\sigma}^{-} - {\sigma'}^{-} ]^{3/2} [dX^{+}][dX^{-}][dX^{i}] 
\nonumber \\ 
& & [du][dv][dw_{i}] [dP^{-}][dP^{+}][dP_{i}] 
\nonumber \\
& & [dp_{u}][dp_{v}] [dp_{w_{i}}] \delta [(P^{+} + \frac{T}{2} {\partial}_{-} X^{+} ) \approx 0] 
\nonumber \\
& & \delta [(P^{-} + \frac{T}{2}  {\partial}_{-} X^{-}) \approx 0] \delta [ ( P_{i} + \frac{T}{2}  {\partial}_{-} X^{i} ) \approx 0].
\label{11}
\end{eqnarray}

The LC Hamiltonian and path integral quantization of the Polyakov D1 brane action $S^{N} $ under the conformal gauge using the ELCWST framework on the hyperplanes of the world-sheet defined by LC world-sheet time: $ {\sigma}^{+} = ( \sigma + \tau )$ = constant is now complete. Also because this is a (conformally) gauge-fixed action, the theory is therefore gauge noninvariant as expected and the associated constraints of the theory form a set of second-class constraints. The problem of operator ordering that occurs while making a transition from the Dirac brackets to the corresponding commutation relations could be resolved  by demanding that all the string fields and momenta of the theory are Hermitian operators and that all the canonical commutation relations be consistent with the hermiticity of these operators \cite{2}. It is important to mention here that in our work we have not imposed any boundary conditions (BC's) for the open and closed strings separately. There are two ways to take these BC's into account: (a) one way is to impose them directly in the usual way for the open and closed strings separately in an appropriate manner \cite{1,2}, and (b) an alternative second way is to treat these BC's as the Dirac primary constraints \cite{6} and study the theory accordingly \cite{6}. In conclusion, the conformally gauge-fixed Polyakov D1 brane action is seen to be a constrained system in the sense of Dirac when it is considered on the light-front in contrast to the case when it is consdired in the instant-form. The model is quantized using the standard constraint quantization techniques on the light-front.

\end{document}